\documentstyle[12pt]{article}
\begin{document}
%%%%%%%%%%%%%%%%%%%%%%%%%%%%%%%%%%%%%%%%%%%%%%%%
%%%%%%%%%%%%%%%%%%%%%%%%%%%%%%%%%%%%%%%%%%%%%%%%
\newcommand{\beq}{\begin{equation}}
\newcommand{\eeq}{\end{equation}}
\newcommand{\ba}{\begin{array}}
\newcommand{\ea}{\end{array}}
\newcommand{\beqa}{\begin{eqnarray}}
\newcommand{\eeqa}{\end{eqnarray}}
\newcommand{\bd}[1]{ \mbox{\boldmath $#1$}  }
\newcommand{\ct}[1]{$^{\mbox{\scriptsize\cite{#1}}}$}
\newcommand{\tresj}[6]{\left(\!\!\!
                            \begin{array}{ccc}
                                      #1&#2&#3\\
                                      #4&#5&#6
                            \end{array}\!\!\!
                       \right)}
\newcommand{\seisj}[6]{\left\{\!\!\!
                            \begin{array}{ccc}
                                      #1&#2&#3\\
                                      #4&#5&#6
                            \end{array}\!\!\!
                       \right\}}
\newcommand{\nuevej}[9]{\left\{\!\!\!
                            \begin{array}{ccc}
                                      #1&#2&#3\\
                                      #4&#5&#6\\
                                      #7&#8&#9
                            \end{array}\!\!\!
                       \right\}}
%%%%%%%%%%%%%%%%%%%%%%%%%%%%%%%%%%%%%%%%%%%%%%%%
%%%%%%%%%%%%%%%%%%%%%%%%%%%%%%%%%%%%%%%%%%%%%%%%
\title{Probing Deformed Orbitals with
       $\vec{\mbox{A}}(\vec{e}, e' N)$B reactions}
\author{ \protect\hspace{-7mm}
  E. Garrido, J.A. Caballero, E. Moya de Guerra, P. Sarriguren, and
\\
J.M. Ud\'{\i}as\thanks{Present address: National Laboratory for Nuclear and
High--Energy Physics, Section K (NIKHEF--K), P.O. Box 41882, NL--1009 DB
Amsterdam, The Netherlands}}
\maketitle
\begin{center}
Instituto de Estructura de la Materia, Consejo Superior de
Investigaciones Cient\'{\i}ficas,\\ Serrano 119, 28006 Madrid, Spain
\end{center}

\thispagestyle{empty}
%\newpage
\begin{abstract}
We present results for response functions and asymmetries
in the nuclear reactions $^{37}\vec{\hbox{Ar}}(\vec{e},
e' n)^{36}\hbox{Ar}$ and
$^{37}\vec{\hbox{K}}(\vec{e},e' p)^{36}\hbox{Ar}$ at
quasifree kinematics. We compare PWIA results obtained using deformed
HF wave functions with PWIA and DWIA results obtained assuming a spherical
mean field. We show that the complex structure of the deformed orbitals can
be probed by coincidence measurements with polarized beam and
targets.
\end{abstract}
\newpage
\section{Introduction}
\indent

The concepts of nuclear deformation and of deformed mean field potential
have been extremely successful, and have allowed to make enormous
progress in our understanding of nuclei ever after the pioneering works of
Rainwater\ct{rain}, Bohr and Mottelson\ct{bm}, and Nilsson\ct{nil}. A basic
notion that comes along
with those concepts is that, for open shell nuclei, single--nucleon
states are not eigenstates of angular momentum, or in short, that
nucleons move in deformed orbitals. In spite of the rich
phenomenology, too vast to be quoted here, up to
date understood on the basis of these concepts, deformed
orbitals have not yet been ``seen". While spherical orbitals have clearly
been seen in electron induced one nucleon knock--out reactions\ct{mlw},
there is no such a direct evidence of the existence of deformed orbitals.

In a spherical nucleus, each single--particle momentum distribution
$n^{ \{n \ell j \} }(p)$ is characteristic of the spherical
orbital $\{n \ell j \}$ occupied
by the nucleon. Single--particle momentum distributions are probed in one
nucleon knock--out reactions and, as already indicated, in the past years
very precise coincidence $(e, e' N)$ measurements have made possible to
map out the momentum distributions of various spherical bound
orbitals\ct{mlw,mas}. An interesting question is whether such experiments
can also measure momentum distributions of deformed orbitals.

In a deformed nucleus the momentum distribution of a given deformed
orbital $i$ is in general a linear combination of spherical orbitals
($n^i(p) = \sum_{\ell j} c_{\ell j}^i n_{\ell j}(p)$). The linear
combination depends on the deformation of the mean field and on the
particular orbital $i$. Thus, for each orbital $i$ and mean field
deformation there is a characteristic momentum distribution
that can in principle be identified by these reactions. Indeed, some of
the existing experimental data on $(e, e' p)$ from $^{28}$Si and
$^{146}$Nd can be explained on the basis of the deformed
model\ct{si28,nd146}. In particular, a comparison of data on quasielastic
electron scattering from $^{142}$Nd and from $^{146}$Nd\ct{nd146,amalf}
confirmed the presence of deformation effects that had been discussed in
Refs.~\cite{ddhf,nikhef}. As pointed out in these references, for
even--even targets the main effects of deformation in $(e, e' p)$ are
related to the higher level density of low--lying states in the residual
nucleus and to the fragmentation of strength of spherical orbitals.
With even--even targets it is not possible to measure the total
momentum distribution of the deformed orbital in a single transition,
since for transitions to a well resolved discrete state of
the residual nucleus only a given component $\ell j$ of the deformed state
$i$ enters. Therefore the interpretation of data is not free of
ambiguities. In this work we show that the most direct evidence for the
existence of deformed orbitals can be gained by using deformed odd--A
targets that can be polarized.

Recently we have investigated spin--dependent momentum distributions in
odd--A deformed nuclei and spectral functions for quasielastic electron
scattering with polarized beam and target\ct{annals,nucph}. We have found
that there are new features of nuclear structure and spin degrees of
freedom that can be revealed in these reactions, and deserve further study.
This paper is devoted to study the new features connected with the role of
$\ell$--wave mixing in deformed single--particle orbitals.

We show that measuring response functions and/or asymmetries in
$\vec{\mbox{A}}(\vec{e},e' p)$B reactions for specific transitions in
deformed nuclei one can see the $\ell$--wave mixing in
deformed orbitals. To this end we show results for response functions and
asymmetries for $^{37}\vec{\mbox{K}}(\vec{e},e'
p)^{36}\mbox{Ar}$  and for $^{37}\vec{\mbox{Ar}}(\vec{e},e'
n)^{36}\mbox{Ar}$, where the residual nucleus is left in its first
excited state. The target and residual nuclei are described within the
scheme of the rotational model of Bohr and Mottelson\ct{bm2}, using as
intrinsic wave functions the Slater determinants obtained from density
dependent Hartree--Fock (DDHF) calculations.
We show the results obtained from DDHF calculations, and stress the role
of nuclear deformation and $\ell$-wave
mixing by comparison to results obtained in the spherical limit.

Most of the results presented here are based on the plane wave impulse
approximation (PWIA) that neglects final state interactions (FSI), allowing
to identify nuclear structure effects in a more transparent way.
We also present a discussion of results obtained in DWIA (distorted wave
impulse approximation) to illustrate
the fact that the effect of deformation discussed here is quite different
from standard FSI effects, and cannot be masked by them.

This paper has the following structure. In section II we give a brief
summary of theory and details of calculations. In section III we show and
discuss the results on response functions and polarization observables
for $^{37}\vec{\mbox{K}}(\vec{e},e'p)^{36}\mbox{Ar}$ and
$^{37}\vec{\mbox{Ar}}(\vec{e},e'n)^{36}\mbox{Ar}$, obtained in PWIA
with the deformed (DDHF) wave functions. In section IV we show the
corresponding results on response functions and asymmetries obtained in
the spherical limit and compare PWIA and DWIA results. Our final remarks
are summarized in section V.

\section{Brief summary of theory and details of calculations}

\subsection{$\vec{\mbox{A}}(\vec{e},e' p)$B reactions}
\indent

The general formalism of quasielastic scattering from complex nuclei with
polarized beam and target has been presented in several
works\ct{nucph}$^,$\ct{don}$^-$\ct{otr}. Here we follow the conventions and
notations of Donnelly and collaborators\ct{nucph,don,dp} and summarize
briefly the equations of interest to describe processes of the type
$\vec{\mbox{A}}(\vec{e},e' N)$B for transitions to discrete states
$J_B^\pi$ in the residual nucleus. Here we mainly
reformulate some of the equations given in past works to make the
formalism more suitable to discuss the points of interest in this paper.

We use kinematical variables in the laboratory frame as follows:
$k^\mu=(\epsilon,\bd{k})$ and $k^{\prime \mu}=(\epsilon',\bd{k}')$ are
initial and final electron four--momenta; $P_A^\mu=(M_A,\mbox{\bf 0})$,
$P_B^\mu=(E_B,\bd{P}_B)$ and $p_N^\mu=(E_N,\bd{p}_N)$ are the
four--momenta of the target nucleus, residual nucleus, and outgoing
nucleon,
respectively; $Q^\mu=(\omega,\bd{q})=k^\mu-k^{\prime
\mu}=p_N^\mu+P_B^\mu-P_A^\mu$ is the four--momentum transferred by the
virtual photon. The laboratory reference frame $(x,y,z)$  is defined by
the unit vectors
\beq
\bd{e}_z=\frac{\bd{q}}{q}
\hspace{1cm}
\bd{e}_y=\frac{\bd{k}_i \times \bd{k}_f}{k_i k_f \sin{\theta_e}}
\hspace{1cm}
\bd{e}_x=\bd{e}_y \times \bd{e}_z
\eeq

Following Refs.~\cite{nucph,don} we write the differential cross section
for the $\vec{\mbox{A}}(\vec{e},e' p)$B process as
\beq
\frac{d\sigma}{d\epsilon' d\Omega_e d\Omega_N}=
\Sigma+ h \Delta =
\Sigma_0\left[ 1 + P_\Sigma + h P_\Delta  \right]
\label{1}
\eeq
where $h$ is the helicity of the longitudinally polarized electron, and
$\Sigma_0$ is the differential cross
section for unpolarized beam and target. $P_\Sigma$ depends on the target
polarization and $P_\Delta$ contributes for polarized beam and
target. Note that for unpolarized target $P_\Delta$ is zero in PWIA, but
in general it is not necessarily zero\ct{otr}$^-$\ct{last}.
Using current conservation to separate transverse and
longitudinal
components of the electron and nuclear currents, the following
standard expressions are obtained for the helicity sum
($\Sigma$) and the helicity difference ($\Delta$) cross sections
\beq
\Sigma={\cal K} \sum_X V_X R^X  \hspace{2cm}
    X=L,T,LT,TT
\eeq
\beq
\Delta={\cal K} \sum_{X'} V_{X'} R^{X'}  \hspace{2cm}
    X'=LT',T'
\eeq
where $X$ runs over the four response functions longitudinal
(L), transverse (T), transverse--longitudinal (LT) and
transverse--transverse (TT) interferences, and $X'$ runs over the other
two possible interference response functions, transverse--longitudinal
(LT') and transverse--transverse (T'). The kinematical factor ${\cal K}$
is given by
\beq
{\cal K}= \sigma_{\mbox{\scriptsize M}} \,
          \frac{p_N M_N M_B}{(2 \pi)^3 M_A f_{\mbox{\scriptsize rec}}},
\eeq
with $\sigma_{\mbox{\scriptsize M}}$ and $f_{\mbox{\scriptsize rec}}$ the
standard Mott cross
section and recoil factor, respectively. Explicit expressions for the other
kinematical factors  $V_X$ and $V_{X'}$ can be found in
Ref.~\cite{don}. Here we only recall that these kinematical factors
have different dependences on the electron kinematics, which together with
the various dependences on polarization and outgoing nucleon directions,
allow in principle to separate the different response functions.

In PWIA, each of the hadronic response functions $R^X$, $R^{X'}$ can in turn be
factored into a part that depends on the nuclear structure and a part that
depends on the half--off--shell nucleon current:
\beqa
R^X&=& {\cal R}^X \overline{M}^{(J_A \rightarrow J_B)}(\bd{p},\bd{P}^\ast)
           \label{4}        \\
R^{X'}&=&\bd{\cal R}^{X'} \bd{\cdot}
      \widehat{\bd{M}}^{(J_A \rightarrow J_B)}(\bd{p},\bd{P}^\ast)
           \label{5}
\eeqa

The single--nucleon responses ${\cal R}^X$, $\bd{\cal R}^{X'}$ have been
studied in Ref.~\cite{dp} for two different choices (cc1 and cc2) of the
half--off--shell nucleon current operator, as well as for different
prescriptions concerning current conservation. For instance, once cc1 is
chosen one may impose current conservation and eliminate the third
component of the single--nucleon current in favour of the zero component.
This prescription is called cc1$^{(0)}$. On the contrary, one may
eliminate the zero component in favour of the third (cc1$^{(3)}$), or one
may simply calculate independently the four components without imposing
current conservation (ncc1). Of course for a free nucleon the same result
is obtained with any of these prescriptions, but in quasielastic
scattering the initial nucleon is not on shell, and different
prescriptions may lead to different results. We shall come back to this
point in section II.3.

The dependence on the nuclear structure in
eqs.~(\ref{4}) and (\ref{5}) is contained in the scalar and
vector momentum distributions $\overline{M}$ and $\widehat{\bd{M}}$, that
depend not only on the momentum
$\bd{p}$ ($=-\bd{P}_B$) of the struck nucleon but also on the target
polarization $\bd{P}^\ast$, and on the particular nuclear transition
($J_A \rightarrow J_B$) under consideration. The scalar and vector
momentum distributions are defined in terms of the spin dependent momentum
distribution matrix
$M_{\sigma \sigma'}^{(J_A \rightarrow J_B)}(\bd{p}, \bd{P}^\ast)$ as
follows\ct{annals}
\beqa
\overline{M}^{(J_A \rightarrow J_B)}(\bd{p}, \bd{P}^\ast)&=&
Tr \left[ M^{(J_A \rightarrow J_B)}(\bd{p}, \bd{P}^\ast)   \right]
                         \\
\widehat{\bd{M}}^{(J_A \rightarrow J_B)}(\bd{p},\bd{P}^\ast)&=&
Tr \left[\bd{\sigma}
   M^{(J_A \rightarrow J_B)}(\bd{p}, \bd{P}^\ast)      \right]
        \label{new9}
\eeqa
where traces are taken in spin space, and the matrix
$M_{\sigma \sigma'}^{(J_A \rightarrow J_B)}(\bd{p},
\bd{P}^\ast)$ is related to the spin dependent polarized spectral
function defined in Ref.~\cite{nucph} by
\beq
S_{\sigma \sigma'}(\bd{p}, E;\bd{P}^\ast)=
  \sum_{J_B} \delta(E-M_A^0+E_{J_B})
  M_{\sigma \sigma'}^{(J_A \rightarrow J_B)}(\bd{p},\bd{P}^\ast)
\eeq
or
\beq
  M_{\sigma \sigma'}^{(J_A \rightarrow J_B)}(\bd{p},\bd{P}^\ast)=
\sum_{M_B}\langle J_A J_A(\Omega^\ast)|
  a_{\mbox{\scriptsize\bf{p}} \sigma}^+ |J_B M_B \rangle
\langle J_B M_B|a_{\mbox{\scriptsize\bf{p}} \sigma^\prime}
    |J_A J_A(\Omega^\ast) \rangle
\eeq

Here and in what follows we assume for simplicity that the target is 100\%
polarized in the direction $\bd{P}^\ast$, characterized by the angles
$\Omega^\ast$. Hence the target has angular
momentum projection $M_A=J_A$ in this direction. $\sigma$ and
$\sigma^\prime$ denote spin projections of the struck nucleon on the
laboratory $z$--axis.

To match with the notation in Refs.~\cite{nucph,don,dp} we denote by $l$
(longitudinal) the vector components of $\bd{\cal R}^{X'}$ and
$\widehat{\bd{M}}^{(J_A\rightarrow J_B)}$ along $\bd{q}$ ($z$--axis), and
by $s$ (side--ways), and $n$ (normal) the vector components in the
perpendicular ($x,y$) plane.

 From eqs.~(\ref{1}) to (\ref{new9}) it can be seen that the vector
momentum distribution
only contributes for polarized beam and target. Furthermore, in
Ref.~\cite{annals} it was shown that this new momentum distribution depends
only on the odd--nucleon wave function and can be probed in transitions to
the ground state rotational band of the residual nucleus. Therefore we
focus on transitions to states $J_B$ in the ground state band
($K_B^\pi=0^+$) of the residual nucleus.

General expressions for spectral functions and momentum distributions for
arbitrary target polarization can be found in Refs.~\cite{annals,nucph}.
Most of the results shown here are for 100\% polarized target in the
$z$--direction ($\bd{P}^\ast \parallel \bd{q}$).
In this particular case, the scalar and vector
momentum distributions are given by
\beq
\overline{M}^{(J_A \rightarrow J_B)}=\sum_{\lambda \ \mbox{\scriptsize even}}
P_\lambda(\cos \theta_p)
\sum_{\ell j \ell' j'} {\cal F}^\lambda(\ell j;\ell' j') \,\,
\tilde{\phi}_{\ell j}(p) \tilde{\phi}_{\ell' j'}(p)
\label{10}
\eeq
\beqa
\widehat{M}_s^{(J_A \rightarrow J_B)}&=&
            \cos\phi_p \, \widehat{M}_t^{(J_A \rightarrow J_B)}  \\
\widehat{M}_n^{(J_A \rightarrow J_B)}&=&
            \sin\phi_p \, \widehat{M}_t^{(J_A \rightarrow J_B)}
\eeqa
\beq
\widehat{M}_{
        \left[ \ba{l}
           \!\!l \!\! \\
           \!\!t \!\!  \ea  \right]   }^{(J_A \rightarrow J_B)} =
  \sum_{L \ \mbox{\scriptsize even}}
    \left[  \ba{l}
             P_L(\cos \theta_p) \\
             P_L^1(\cos \theta_p)  \ea \right]
 \sum_{\ell j \ell' j'}
  {\cal G}^L_{
        \left[  \ba{l}
           \!\!l \!\! \\
           \!\!t \!\!  \ea  \right]   }(\ell j; \ell' j') \,\,
  \tilde{\phi}_{\ell j}(p) \tilde{\phi}_{\ell' j'}(p)
\label{13}
\eeq
with $p$, ($\theta_p, \phi_p$) the absolute value and direction of the
momentum $\bd{p}$ of the bound nucleon. $\tilde{\phi}_{\ell j}(p)$ and
$\tilde{\phi}_{\ell' j'}(p)$ are the spherical wave projections of the
deformed odd--nucleon wave function in momentum space (see section II.2),
and the coefficients ${\cal F}$, ${\cal G}$ are given by:
\beqa
{\cal F}^\lambda(\ell j;\ell' j')&=&\frac{\hat{\lambda}^2 \hat{J}_A^2
 \hat{J}_B^2 \hat{\ell} \hat{\ell}' \hat{j} \hat{j}'}{2 \pi}
 (-1)^{J_A+j+j'-\frac{1}{2}} \tresj{J_A}{J_A}{\lambda}{-J_A}{J_A}{0}
 \tresj{J_B}{J_A}{j}{0}{K_A}{-K_A}
 \nonumber \\
&&
 \tresj{J_B}{J_A}{j'}{0}{K_A}{-K_A}
 \tresj{\ell}{\ell'}{\lambda}{0}{0}{0}
 \seisj{j'}{j}{\lambda}{\ell}{\ell'}{\frac{1}{2}}
 \seisj{j'}{j}{\lambda}{J_A}{J_A}{J_B}
     \label{calf}
\eeqa
\beqa
  {\cal G}^L_{
        \left[  \ba{l}
           \!\!l \!\! \\
           \!\!t \!\!  \ea  \right]   }(\ell j; \ell' j')&=&
-\frac{\sqrt{3}}{\pi} \hat{L}^2 \hat{J}_A^2 \hat{J}_B^2 \hat{\ell}
 \hat{\ell}' \hat{j} \hat{j}' (-1)^{J_A+\ell'+j}
\sum_{\lambda=L \pm 1} \hat{\lambda}^2
\tresj{J_A}{J_A}{\lambda}{-J_A}{J_A}{0} \nonumber \\
&&\tresj{J_B}{J_A}{j}{0}{K_A}{-K_A}
\tresj{J_B}{J_A}{j'}{0}{K_A}{-K_A}
\tresj{\ell}{\ell'}{L}{0}{0}{0}
\nuevej{L}{\lambda}{1}{\ell'}{j'}{\frac{1}{2}}{\ell}{j}{\frac{1}{2}}
\nonumber \\
&&\seisj{j'}{j}{\lambda}{J_A}{J_A}{J_B}
\left[
     \ba{l}
   -\frac{1}{\sqrt{2}}\tresj{1}{L}{\lambda}{0}{0}{0} \\
    \frac{1}{\sqrt{L(L+1)}}\tresj{1}{L}{\lambda}{-1}{1}{0}  \ea \right]
\label{15}
\eeqa

We would like to remark that the sum over $\ell j, \ell' j'$ is
characteristic of nucleon knock--out from a deformed orbital. When the
struck nucleon is in a spherical orbital $\{n \ell j \}$ this sum reduces
to a single term.

It is easy to check that for $J_B=0$
({\it i.e.}, for transitions to the ground state of the residual nucleus) the
coefficients ${\cal F}^\lambda(\ell j; \ell' j')$ and
${\cal G}^L_{l,t}(\ell j; \ell' j')$ are zero unless $j=j'=J_A$. This
means that for transitions to the ground state in the residual nucleus
only one component ($\ell j = \ell' j'$) contributes. For the nuclei under
consideration only the $d_\frac{3}{2}$ component would contribute. Hence
this transition ($J_A^\pi=\frac{3}{2}^+ \rightarrow J_B^\pi=0^+$) allows us
to see only one $\ell j$--component of the deformed orbital,
and it is more interesting to consider the transition to the first excited
state ($2^+$) that, as we shall see, allows us to explore the deformed
orbital in its full complexity. In what follows we focus on the transition
$J_A^\pi=\frac{3}{2}^+ \rightarrow J_B^\pi=2^+$.

As it is discussed in the next section, for the two target nuclei
considered here the ground state rotational band has $J_A=\frac{3}{2}$ and
$K_A=\frac{1}{2}$. The coefficients ${\cal F}$ and ${\cal G}$ for these
$J_A$, $K_A$ values and for $J_B=2$ are given in tables I--III.

For unpolarized target only the $\lambda=0$ multipole of the scalar
momentum distribution ($\overline{M}_{\lambda=0}^{(J_A \rightarrow J_B)}$)
contributes to the differential cross section $\Sigma_0$.

The notation for momentum distributions used here follows that of
Ref.~\cite{annals}, and is closely related to that used in Ref.~\cite{nucph}
were we used the notation $N_0^{J_B}$, $N_l^{J_B}$, $N_s^{J_B}$ and
$N_n^{J_B}$ in place of $\overline{M}^{(J_A \rightarrow J_B)}$,
$\widehat{M}^{(J_A \rightarrow J_B)}_l$,
$\widehat{M}^{(J_A \rightarrow J_B)}_s$ and
$\widehat{M}^{(J_A \rightarrow J_B)}_n$, respectively. We would like to
point out that
the expressions given in appendix B of Ref.~\cite{nucph} are
for arbitrary target polarization and polarization direction, as well as
for arbitrary $J_A \rightarrow J_B$ transition, while the ones given in
eqs.~(\ref{10}) to (\ref{15}) are for a 100\% polarized target in the $z$
direction and for a transition to the state $J_B$ in the ground state
rotational band of the residual nucleus.

\subsection{Nuclear structure calculation}
\indent

The scalar and vector momentum distributions defined in the previous
section have been calculated with the wave functions obtained
from DDHF calculations, assuming a time reversal invariant axially
symmetric mean field. As in previous work on momentum distributions in
deformed nuclei\ct{ddhf,annals,nucph} we have used the McMaster\ct{mac}
version of the HF program, and the SKA interaction\ct{koh}. The targets
and residual nuclei are described by the same mean field. We have ignored
pairing correlations that cannot be reliably handled by BCS approximation
in these $sd$ shell nuclei. The calculated self--consistent  mean field is
oblate ($Q_0 < 0$) and the odd nucleon (a neutron in $^{37}$Ar and a proton
in $^{37}$K) is in a $K^\pi=\frac{1}{2}^+$ state, and has basically the
same wave function in both cases. This can be seen in table IV, where we
give the weights of the $\ell j$--projection of the odd--nucleon wave
function in both cases.

In table V we show the theoretical results
obtained for the ground state binding energies, charge r.m.s radii, and
magnetic moments, as well as for intrinsic quadrupole moments, moments of
inertia and decoupling parameters. The properties of the rotational bands
are calculated according to the expressions given in Ref.~\cite{pr}, in
particular, we use the cranking formula for the moments of inertia.
The available experimental data are also given in the table.
It is also worth pointing out that data on r.m.s. radii\ct{dat} and
quadrupole moments\ct{led,prom} for other neighbour nuclei like $^{37}$Cl
and $^{39}$K are close to the theoretical results in table V. In
particular, the negative values of the intrinsic quadrupole moments are in
good agreement with the experimental\ct{led,prom} quadrupole moment for
$^{36}$Ar ($Q_0^{\mbox{\scriptsize exp}}=-38.5 \pm 21.0$).

As seen in table V we get good agreement with experiment for binding
energies and magnetic moments. Also, the experimental spectral
sequence\ct{led} $\left( \frac{3}{2}^+ \right.$, $\frac{1}{2}^+$,
$\frac{7}{2}^+$,
$\left. \frac{5}{2}^+\right)$ of angular momentum states in $^{37}$Ar and
$^{37}$K is
well reproduced with the moments of inertia and decoupling parameters in
table V, although the first excited state ($\frac{1}{2}^+$) is lower than
the experimental $\frac{1}{2}^+$ state. Actually, the $\frac{7}{2}^+$
states that we get at $E\sim 2.6$ MeV in both $^{37}$Ar and $^{37}$K,
correspond quite nicely with the experimental $\frac{7}{2}^+$ state at
2.22 MeV in $^{37}$Ar and
with a state at 2.28 MeV in $^{37}$K whose spin and parity have not yet
been assigned. Also the $\frac{5}{2}^+$ states at 2.80 and 2.75 MeV in
$^{37}$Ar and $^{37}$K, respectively, are fairly well reproduced
theoretically (with the ${\cal I}$ and $a$ values in table V we get
$E\sim 2.85$ and 2.86 MeV, respectively). The $2^+_1$ state in $^{36}$Ar is
also close to the one we get with the moment of inertia
${\cal I}^{\mbox{\scriptsize th.}}=2.5$ MeV$^{-1}$. In summary, our DDHF
calculations give a fairly accurate description of the properties of the
ground state bands of the
nuclei involved in the transitions under study.

Coming back to the calculations of momentum distributions, as seen from
eqs.~(\ref{10}) to (\ref{13}), for the transitions considered all we need
is to compute the spherical $\ell j$--wave projection of the deformed
odd--nucleon wave function in momentum space. The latter are obtained as
follows. From the Hartree--Fock output we get the odd--nucleon wave
function in coordinate space and cylindrical coordinates with intrinsic
angular momentum projection $K=\frac{1}{2}$ and parity $\pi=+1$:
\beq
\Phi({\bd{r},s,q})=\chi_\tau(q) \sum_\alpha C_\alpha
\varphi_\alpha(\bd{r},s)
\label{16}
\eeq
where $\chi_\tau(q)$ denotes isospin function and where $\alpha$ denotes
the set of quantum numbers $\alpha=[n_r,n_z,\Lambda,\Sigma]$ with
$\Lambda+\Sigma=K=\frac{1}{2}$, $\, \Lambda \geq 0$ and $2n_r+n_z+\Lambda=N=$
even. The sum extends over all basis states from $N=0$ to $N=10$.
$C_\alpha$ are the HF coefficients and $\varphi_\alpha$ are the basis
wave functions\ct{vaut} normalized to 1.

Taking the Fourier transform in eq.~(\ref{16}), we get the odd--nucleon wave
function in momentum space
\beq
\tilde{\Phi}(\bd{p},s,q)=\frac{1}{(2 \pi)^\frac{3}{2}}
\int d\bd{r} e^{-i \bd{p} \cdot \bd{r}}
\Phi(\bd{r},s,q)
\eeq
from which the $\ell j$--projection is obtained as
\beq
\tilde{\phi}_{\ell j}(p)=
 \langle {\cal Y}_{\ell j}^{m_j=\frac{1}{2}}(\hat{\bd{p}},s)|
 \tilde{\Phi}(\bd{p},s) \rangle
\eeq
where the brackets indicate integration over $\bd{p}$--direction and
spin, and where
isospin labels and functions have been omitted to simplify the notation.
Using the standard definition of the vector spherical harmonics
\beq
{\cal Y}_{\ell j}^{m_j}(\hat{\bd{p}},s)=
\sum_{m \sigma}\langle \ell m \frac{1}{2} \sigma | j m_j \rangle
    Y_{\ell}^m(\hat{\bd{p}}) \chi_\sigma(s)
\eeq
we get
\beq
\tilde{\phi}_{\ell j}(p)=\sum_\alpha C_\alpha
\langle \ell \Lambda \frac{1}{2} \Sigma |j \frac{1}{2} \rangle
R_\ell^\alpha(p)
\eeq
with
\beqa
R^{\alpha}_{\ell}(p)&=& (-i)^{N+2\Lambda}\sqrt{\pi_i}
	\left[\frac{(2\ell+1)(\ell -\Lambda)!}{2(\ell +\Lambda)!}\right]
		^\frac{1}{2}
	\nonumber \\
& & 	\int_0^{\pi}\sin\theta_p \, d\theta_p \,
          P_\ell ^\Lambda (\cos\theta_p) \,
	\psi_{n_\perp}^\Lambda(p \sin\theta_p) \, \psi_{n_z}(p\cos \theta_p)
	\nonumber \\
 &  &
\eeqa
and
\beq
\psi_{n_\perp}^{\Lambda}(p_{\perp})=
		\left[
			\frac{2n_\perp !}{\beta^2_\perp (n_\perp +\Lambda)!}
		\right]^\frac{1}{2}
			\eta^{\Lambda/2}e^{-\eta/2}L^{\Lambda}_{n_\perp}(\eta) ,
\eeq

\beq
\psi_{n_z}(p_z)=
		\left[
			\frac{1}{\beta_z \sqrt{\pi} 2^{n_z}n_z !}
		\right]^\frac{1}{2}
			e^{-\xi^2 /2}H_{n_z}(\xi)
\eeq
where $L_{n_\perp}^\Lambda$ and $H_{n_z}$ are associated Laguerre and
Hermite polynomials respectively, $\eta=p_\perp^2 / \beta_\perp^2$,
$\xi=p_z / \beta_z$, and $\beta_\perp$, $\beta_z$ are the inverse of
the harmonic oscillator length parameters
($\beta_{\perp,z}=\sqrt{m \omega_{\perp,z}}$).
(Natural units $c=\hbar=1$ have been used throughout).

As seen in table IV for both $^{37}$K and $^{37}$Ar the dominant $\ell
j$-wave component is the $d_\frac{3}{2}$ ($\sim 75\%$), but the
$s_\frac{1}{2}$ component is also important ($\sim 19 \%$). Of decreasing
importance are the higher $\ell j$ components $d_\frac{5}{2}$ ($\sim 5\%$)
and $g_\frac{7}{2}$ ($\sim 0.7 \%$), while even higher components contribute
less than $0.1 \%$.

\subsection{Choice of kinematics and single--nucleon \protect\\ responses}
\indent

We are interested in coincidence measurements in the quasifree region
and in transitions to the lower states in the residual nucleus to probe
essentially the odd--nucleon wave function. Our interest is to choose the
kinematics in such a way that the kinetic energy of the outgoing nucleon
is large
enough to minimize effects of final state interactions, and yet the cross
sections are as large as possible.
Though for comparison with data, FSI have to be taken into account (see for
instance Ref.~\cite{mlw}), at sufficiently high momentum transfer,
and for an energy transfer at the quasielastic peak, the process is
expected to be less influenced by final state interactions and exchange
effects. Hence we select the kinematics by fixing the value of the
momentum transfer $q$ to be reasonably large below 1 GeV. For the
selected $q$ value, we choose the energy transfer $\omega$ to be that
at the quasielastic peak
\beq
\omega=\omega_{\mbox{\scriptsize q.p.}}=
\sqrt{q^2+M^2}-M+E_s
\eeq
where $E_s$ is the neutron (proton) separation energy in $^{37}$Ar
($^{37}$K). Solving the energy balance equation
\beq
\sqrt{p_N^2+M^2}+\sqrt{p^2+M_B^2}=M_A+\omega
\eeq
for $p_N=|\bd{q}-\bd{p}|$, and subject to the condition $-1\leq
\cos\theta_p \leq 1$, one can obtain the range of allowed values  of the
bound nucleon momentum $p_{\mbox{\scriptsize min}} \leq p \leq
p_{\mbox{\scriptsize max}}$. For fixed values of $q$ and $\omega$ as well
as fixed mass $M_B$ of the residual nucleus one can vary the bound nucleon
momentum $p$ in the interval between $p_{\mbox{\scriptsize min}}$ and
$p_{\mbox{\scriptsize max}}$ by varying the angle $\theta_N$. For the
cases of interest here in which the mass of the residual nucleus is close
to the ground state mass ($M_B = 33.5$ GeV $\gg q$) one
has that the allowed range of variation for $p$ is between
$p_{\mbox{\scriptsize min}}=0$ and $p_{\mbox{\scriptsize max}}\sim 2q$.

Of course as $\theta_N$ varies not only $p$ varies, but also $p_N$ has to
change to obey energy--momentum conservation. However, from the
expression for $p_N$
\beq
p_N=p_N(p)=\left[
  \left( \omega+M_A-\sqrt{p^2+M_B^2} \right)^2-M^2
           \right]^\frac{1}{2}
\eeq
one sees that the variation of $p_N$ is negligible, since one has that
$M_B^2 \gg p^2$ in the whole range of variation of $p$. The maximum value
of $p_N$ ($p_N=q$ for $p=0$) decreases at most $1.5\%$ in going from
$p=p_{\mbox{\scriptsize min}}$ to $p=p_{\mbox{\scriptsize max}}$ for
$q=500$ MeV. Hence FSI effects that depend on $p_N$ remain essentially
constant for this kinematics. It should be remarked that in this
kinematics $\theta_p$ and $p$ are no longer independent but are related by
\beq
\cos\theta_p=\cos\theta_p(p)=\frac{p_N^2(p) - q^2 - p^2}{2 p q}
\label{thp}
\eeq

The single--nucleon responses ${\cal R}^X$, $\bd{{\cal R}}^{X'}$ for this
kinematics and assuming $\phi_N=0$ are represented in Figs.~1 and 2,
respectively, for proton (a) and neutron (b) emission and for $q=500$ MeV.
The figures represent the results for the choice cc1$^{(0)}$. At the
kinematics considered, results for cc2$^{(0)}$, ncc1 and ncc2 are similar
to the ones in Figs.~1 and 2. Only the choices cc1$^{(3)}$ and
cc2$^{(3)}$, which are less reliable, give different single--nucleon cross
sections\ct{dp} in the $p$--range considered here.

We have chosen $\phi_N=0$ and hence the normal
components of the $\bd{\cal R}^{X'}$ vectors are zero
(${\cal R}^{LT'}_n={\cal R}^{T'}_n=0$). For $\bd{\cal R}^{LT'}$ the
longitudinal components (${\cal R}_l^{LT'}$) start at zero and
increase  rapidly with increasing $p$ values, while the side--ways
components (${\cal R}_s^{LT'}$) remain practically constant (taking much
larger absolute values in the odd proton case). For $\bd{\cal R}^{T'}$
the opposite is true, the $l$ components remain practically constant
while the $s$ component for the proton increases rapidly starting from
zero (in
the neutron case it remains close to zero in the whole $p$--range).
The scalar response functions ${\cal R}^X$ are all
practically constant with $p$, except ${\cal R}^{LT}$ that starts at zero
and increases rapidly in absolute value with $p$. Of course in the neutron
case only
${\cal R}^T$ and ${\cal R}^{LT}$ take large values, while in the proton
case the large response functions are ${\cal R}^L$, ${\cal R}^T$ and
${\cal R}^{LT}$. The response ${\cal R}^{TT}$ is in all cases very small.
These considerations have to be kept in mind to understand to what extent
the observable response functions ($R^X$ and $R^{X'}$) can reveal nuclear
structure properties. In particular, the fact that ${\cal R}^L$ and
${\cal R}^T$ present a small variation with $p$ justifies the
approximation made in the context of $y$--scaling for inclusive
unpolarized electron scattering\ct{scal}. The strong $p$--dependence of
${\cal R}^{LT}$ indicates that the phenomenon of $y$--scaling may not hold
in more general situations.

The similarity between ${\cal R}^T$ and ${\cal R}^L$ in the proton case
is characteristic of our kinematics. For larger $q$ values, for instance
for $q=800$ MeV, ${\cal R}^T$ is larger than ${\cal R}^L$ in both proton
and neutron cases. Otherwise, at $q=800$ MeV the single--nucleon response
functions show the same qualitative features seen in Figs.~1 and 2, but
quantitatively they are typically 2--3 times smaller, hence giving
rise to smaller cross sections. In next sections all the
results shown are for $q=500$ MeV and for cc1$^{(0)}$.

%%%%%%%%%%%%%%%%%%%%%%%%%%%%%%%%%%%%%%%%%%%%

%%%%%%%%%%%%%%%%%%%%%%%%%%%%%%%%%%%%%%%%%%%%

\section{Response functions and polarization observables}
\indent

In this section we present our results on response functions and polarization
observables and discuss the role of the $\ell$--wave mixing in the deformed
nucleon wave function. To this end we present first in Fig.~3 the scalar
($\overline{M}$) and vector components ($\widehat{M}_l$, $\widehat{M}_t$) of
the
spin dependent momentum distributions defined in
eqs.~(\ref{10})--(\ref{13}) for the
transition from the ground state in $^{37}$K to the $2^+$ state in the ground
state band of $^{36}$Ar. The results for the transition from the ground state
in $^{37}$Ar are similar. The coefficients ${\cal F}$ and ${\cal G}$ for
this transition are
tabulated in tables I--III.

$\overline{M}$, $\widehat{M}_l$ and $\widehat{M}_t$ are
shown in plots (a), (b) and (d) of Fig.~3, respectively. Note that these
momentum
distributions have in principle two independent variables $p$ and $\theta_p$,
however the plots in Fig.~3 correspond to the kinematics discussed in section
II.3 where $\theta_p$ is a function of $p$ (see eq.~(\ref{thp})). In plot (c)
of
Fig.~3 we show for comparison the $\lambda =0$ multipole of the scalar momentum
distribution $\overline{M}_{\lambda=0}$, which is the only one contributing
when the
target is unpolarized. As seen in tables I--III only the diagonal terms
$\ell j=\ell' j'$ contribute to $\overline{M}_{\lambda=0}$, while
$\overline{M}_{\lambda=2}$ and the vector momentum distribution components
$\widehat{M}_l$, $\widehat{M}_t$ contain also important off--diagonal
contributions.
Depending on the $p$ value and on the particular components of the spin
dependent momentum distribution one has dominance of a given $\ell j$ component
or dominance of the interference terms with $\ell j \neq \ell' j'$. It should
be stressed that the interference terms are only present for a polarized
deformed target. In Fig.~3 we also show, for each momentum distribution
component,
the contributions from the most important terms. Looking at tables I--IV it is
easy to find out that in general the main contributions are from the terms with
$\ell=\ell'=0$ ($ss$), $\ell=\ell'=2$ ($dd$), and $\ell=0,\ell'=2$ or
viceversa ($sd$).

As seen in Fig.~3(a) the scalar momentum distribution has a maximum at $p=0$
due to the
$ss$ contribution that dominates in the low $p$ region ($p\leq 0.5$ fm$^{-1}$),
and a second maximum at $p\approx .8$ fm$^{-1}$ that is made up not only from
the $dd$ contribution but also from the contributions of $sd$
and $ss$ terms. Though the scalar momentum distribution in the polarized case
has a more pronounced second maximum than in the unpolarized case (compare
plots (a) and (c) in Fig.~3), the profiles are not very different in the two
cases. On the contrary the vector momentum distribution components $l$, $t$
have
quite different profiles. Particularly, $\widehat{M}_t$ is dominated by the
$sd$ interference term, and therefore contains basically the information
missing
in the unpolarized momentum distribution. The shape of $\widehat{M}_t$ is
characteristic of the $sd$ interference term and its observation would
provide the most direct evidence for the $\ell$--wave mixing in deformed
orbitals.

In Figs.~4 and ~5 we show the response functions for the reactions
$^{37}\vec{\hbox{K}}(\vec{e},e' p)^{36}\hbox{Ar}$ and for
$^{37}\vec{\hbox{Ar}}(\vec{e},e' n)^{36}\hbox{Ar}$,
respectively. Also shown for comparison by dashed lines, are the response
functions
for unpolarized targets. As seen in the figures the profiles of the response
functions reflect quite nicely the features of the momentum distributions
shown in Fig.~3, and depend strongly on the $\ell$--mixing. Clearly, the
relationship between the results in Fig.~4 an those in Fig.~5 is dictated
by the differences in the single--proton and neutron response functions
depicted in Figs.~1 and 2.

The response functions $R^X$ depend on the scalar momentum distribution
and follow closely the profiles in Figs.~3(a) and ~3(c) for the polarized and
unpolarized cases, respectively. This is so particularly for the longitudinal
($R^L$) and transverse ($R^T$) response functions for which
the corresponding single--nucleon responses are fairly constant with $p$
(see Fig.~1). As already remarked, at the kinematics considered here, the
single--proton responses ${\cal R}^L$ and ${\cal R}^T$ are very close and
therefore the total $R^L$, $R^T$ responses in Fig.~4 are practically
identical, while in Fig.~5 $R^T$ is much larger than $R^L$.
The single--nucleon response ${\cal R}^{LT}$ starts at zero
and then goes fast to large negative values. Hence, the response
function $R^{LT}$ produces a quite different mapping than $R^L$ and $R^T$
of the scalar
momentum distribution. The same can be said about $R^{TT}$ which is much
smaller than $R^{LT}$, and, in this sense, is less interesting.

As seen from the comparison to the unpolarized case, the
main information that these responses $R^X$ carry is that corresponding to
the
$ss$ and $dd$ contributions and not much more information seems to be
gained polarizing the target. It is clear that in these response functions
the main observational feature due to deformation is the presence
of the two sizeable peaks, one corresponding to the $s$--wave and the
other corresponding
to the $d$--wave. This feature, which is characteristic of the deformed
orbital, has
already been observed to some extent in quasielastic scattering from deformed
even--even targets\cite{si28,nd146}. The important difference is that for
even--even targets the observation of the effect is associated to lack of
resolution, whereas in the case of odd--A target this feature can be
observed in a transition to a single state in the residual nucleus, as
considered here. Of course,
when the struck nucleon is in a single spherical orbital there is no way to
observe such a pronounced double--peaked structure in a single transition. This
point is illustrated in Section IV where we discuss the spherical limit.

Even more unambiguous evidence on the $\ell$--mixing in the deformed orbitals
can be gained by measuring the response functions for polarized beam and
target,
$R^{LT'}$ and $R^{T'}$. As seen in Figs.~4 and ~5 these response functions
are more
sensitive to the $sd$ interference terms, particularly $R^{LT'}$ whose shape
is basically that of the $sd$ contribution to the transverse vector momentum
distribution (see Fig.~3(d)). $R^{LT'}$ is dominated by the $sd$
contribution while $R^{T'}$ is not. This follows from the facts that
{\it i)} $\widehat{M}_t$ is dominated by the
$sd$ contribution but $\widehat{M}_l$ is not, and {\it ii)} in the $LT'$
response the transverse component of the single--nucleon response (${\cal
R}^{LT'}_s$) dominates over
the longitudinal one (for the odd--neutron case this is only true at low $p$),
while in the $T'$ response the longitudinal single--nucleon response (${\cal
R}^{T'}_l$) dominates.
Thus the new feature of $sd$ interference,
a sign of the deformation of the orbital of the struck
nucleon, can in principle be detected
by measuring the $R^{LT'}$ response at the corresponding kinematics.

It should be emphasized that the choices of kinematics and of polarization
direction
play an important role in dictating which of the response functions ($R^{T'}$
or
$R^{LT'}$, or both) is more sensitive to the $\ell$--wave interference terms.
We have carried out a search for the most favorable conditions to detect such
$sd$--interference terms for the transitions of our concern here. We found that
a particularly favorable case is that in which the target is polarized
perpendicular to the scattering plane ($\bd{P}^\ast \parallel \bd{e}_y$)
and the outgoing nucleon is detected in the plane with $\phi_N \sim
60^\circ$ (remaining kinematics as discussed in section II.3). In Fig.~6
we show the $LT'$ and
$T'$ response functions for this particular case together with the
$ss$, $sd$ and $dd$ contributions. As seen in the figure, in this specific
case the two $R^{X'}$ response functions are large and are dominated by
the $sd$
contribution, hence providing a particularly interesting case for an
experimental test.

We should emphasize however that also for $\bd{P}^\ast \parallel \bd{e}_z$ and
$\phi_N=0$ (the conditions to which Figs.~4 and ~5 apply) the asymmetries
${\cal P}_{\Sigma}$ and ${\cal P}_{\Delta}$ take also
quite sizeable values. As seen in Fig.~7, ${\cal P}_{\Sigma}$ and
${\cal P}_{\Delta}$ oscillate, depending on the $p$--value, between
--0.4 and +0.2 and between --0.1 and +0.5, respectively. These
oscillations depend strongly on the nuclear structure. Actually $P_\Sigma$
is independent of the single--nucleon responses and
depends only on the scalar momentum distribution
\beq
{\cal P}_{\Sigma}=\frac{\Sigma-\Sigma_0}{\Sigma_0}=\frac{\overline{M}}
{\overline{M}_{\lambda=0}}-1.
\label{28}
\eeq

As will be seen in the next section in the spherical limit $P_\Sigma$ is just
proportional to $P_2(\cos\theta_p)$. Hence the strong $p$
dependence observed in the solid lines of Fig.~7 reflects the strong dependence
of ${\cal P}_{\Sigma}$ and ${\cal P}_{\Delta}$ on the complex structure
of the deformed orbital.

%%%%%%%%%%%%%%%%%%%%%%%%%%%%%%%%%%%%%%%%%%%%%%%%%%%%%%%%%%%%%%%%%%%%%%%%%%%%
%%%%%%%%%%%%%%%%%%%%%%%%%%%%%%%%%%%%%%%%%%%%%%%%%%%%%%%%%%%%%%%%%%%%%%%%%%%%

\section{Comparison with spherical limit and considerations on FSI effects}
\indent

To appreciate better the size of the effects of deformation, we compare
here the results obtained in the previous section with the ones obtained
in the spherical limit. It is also important to see whether FSI, that have
not been taken into account up to here, may produce effects in the
response functions and asymmetries comparable to those produced by
deformation.

For these purposes we present in this section PWIA and DWIA
results obtained in the spherical
limit, assuming that the struck nucleon is in a
$d_\frac{3}{2}$ orbital.
The DWIA calculations have been made using a relativistic optical potential
as described in Ref.~\cite{udi}. A discussion of the effects of FSI in
coincidence $(\vec{e},e' p)$ reactions from complex nuclei can be found in
Refs.~\cite{otr,pic},
based on nonrelativistic and relativistic optical potentials, respectively.
To our knowledge, FSI effects have not been studied yet
for the case of $(\vec{e},e' n)$ reactions involving
polarized complex nuclei. This study is complicated by the
fact that optical potentials are not as well determined as they
are in the proton case. Furthermore,
one may have to take into account processes of the
type $(e, e' p)$ followed by $(p,n)$,
especially if the outgoing
neutron is not very energetic ({\it i.e.,\/} when the quasifree
approximation may not be valid and charge--exchange processes may play
a role).  Since the neutron has a very small electric form factor,
such issues can be relevant when the charge/longitudinal
projections of the current are being examined.
An estimate of the contribution of these charge--exchange
processes to the unpolarized $(e,e' n)$
cross section was calculated by Giusti and
Pacati in Ref.~\cite{gius}.
In our case of an odd--neutron target this effect is not expected to be
especially important
as long as we restrict ourselves to missing energies corresponding to the
neutron separation energy, as we do here.
In this case, FSI effects are expected to be comparable to those for
proton knockout, and therefore in this section we show only results for
$^{37}\vec{\mbox{K}}(\vec{e}, e' p)^{36}$Ar.

The results shown in Fig.~8 correspond to the same transition and kinematics
as those shown in Figs.~4--6; except that now we show only the results for
the more representative response functions when the struck nucleon is a proton
($^{37}$K target) in a $d_\frac{3}{2}$ orbital. For $\bd{P}^\ast \parallel
\bd{e}_z$ and $\phi_N=0$ we only show, in addition to the $R^{X'}$
responses, the large $R^X$ responses $R^L$ and $R^{LT}$, because
$R^T$ is very similar to $R^L$ for this
kinematics and $R^{TT}$ is much smaller (see also Fig.~4).
The $R^{X'}$ responses are also shown for $\bd{P}^\ast \parallel
\bd{e}_y$ and $\phi_N=60^\circ$ to compare with the results given in Fig.~6 for
the deformed HF solution.

The effect of deformation stands out clearly when we compare the PWIA
results in Fig.~8 (solid lines) with the corresponding results in Figs.~4
and 6. The complex and varied structures of the responses in Figs.~4 to
6 disappear in the spherical limit. In this limit the shape of all
response functions is similar, and has a single sizeable peak
corresponding to that of
the $d_\frac{3}{2}$ wave function. This dominant feature, as seen in Fig.~8,
prevails in
the DWIA results (dashed lines). The same conclusion is reached if one
assumes that the
struck nucleon was in a single $s_\frac{1}{2}$ orbital (in that case one has in
addition that $\widehat{M}_t=0$ and the $R^{X'}$ responses are tiny in
most cases).

The effect of final state interactions,
taken here into account in the distortion of the outgoing proton wave
function produced
by the nuclear optical potential, varies depending on the kinematics and
on the various response functions, which in some cases can be strongly
suppressed\ct{bm}. This agrees with the analyses carried out in
Ref.~\cite{otr}, where one can also see that FSI effects on specific
response functions can be minimized with appropriate choices of
kinematics, decreasing with increasing energy of the outgoing nucleon. The
FSI effects found here tend to be somewhat larger than those found in
Ref.~\cite{otr} because of the relativistic optical potential used here. A
comparison between calculations with relativistic and nonrelativistic
optical potentials can be found in Refs.~\cite{udi,tesudi}. We do not
enter in details here because the main purpose of the DWIA results shown
in Fig.~8 is to illustrate that typical FSI effects are quite different
from the effects of deformation. As
discussed previously, using the deformed HF solution we found quite
different shapes for the various $R^X$ and $R^{X'}$
responses that we could directly link to the specific role of $\ell$--wave
mixing in each response function.

This conclusion also emerges from the comparison in Fig.~7 between the
results for ${\cal P}_\Sigma$ and
${\cal P}_\Delta$ obtained with the deformed HF solution (solid lines),
and the results obtained in the spherical limit in PWIA (dashed line) and in
DWIA ( short--dashed line). As seen in Fig.~7 the DWIA results obtained
for ${\cal P}_\Sigma$ and ${\cal P}_\Delta$ are very close to the PWIA
results over a fairly extended $p$ range. In this range the effects of FSI
are qualitatively (and in most cases quantitatively) similar in the
polarized and in
the unpolarized responses, and therefore, tend to cancel when calculating
the asymmetries ${\cal P}_\Sigma$ and ${\cal P}_\Delta$. On the contrary, the
effects of deformation are different in the polarized and in the
unpolarized responses, resulting in strong oscillations with $p$ of
${\cal P}_\Sigma$ and ${\cal P}_\Delta$ that are in contrast with the
smooth behaviours observed for the spherical limits. Note that in PWIA
${\cal P}_\Sigma$ is independent of the single--nucleon responses, and that
in the
spherical limit eq.~(\ref{28}) reduces to a second order polynomial in
$\cos^2\theta_p$
\beq
P_\Sigma=\frac{3}{5} \, P_2(\cos{\theta_p}),
\label{ps}
\eeq
with $\theta_p$ given by eq.~(\ref{thp}), which explains the smooth
$p$--dependence of the dashed lines in Fig.~7. When FSI effects are
included, eqs.~(\ref{28}) and (\ref{ps}) do not hold, but differences
between DWIA and PWIA results in the spherical model are notably smaller
than the differences between the PWIA results with deformed and spherical
models. ${\cal P}_\Delta$ depends on the single--nucleon responses even in
PWIA, but in the $p$ region of
interest the dominant responses in the numerator and denominator have a
very similar $p$ dependence when one takes the spherical limit.

Obviously for comparisons to experimental data it would be desirable to
dispose of results
that take into account simultaneously both FSI and deformation effects.
DWIA calculations using deformed HF solutions for the
bound nucleon wave functions, are now in progress. The main difficulty in
this task is
that, for consistency, the existing DWIA codes have to be extended to
take into account deformation in the optical potentials.

Nevertheless the
present results indicate that the effects of deformation encountered here
will prevail and be clearly identifiable when more realistic
calculations beyond PWIA are made.

Thus, although DWIA calculations using the deformed model should be
performed for a quantitative comparison to experimental data, the present
results allow us to conclude that the effects of deformation will not be
masked by those of FSI.

\section{Summary and final remarks}
\indent

The purpose of this work was to study the sensitivity to the $\ell$--wave
mixing in deformed bound orbitals of response
functions and asymmetries in $\vec{\hbox{A}}(\vec{e},e' N)\hbox{B}$
processes. For this purpose we choose to study the transitions from
polarized $^{37}$Ar and $^{37}$K to the first $2^+$ state in
$^{36}$Ar, which we consider to be particularly interesting.
The nuclear structure content of these two transitions is practically
identical, while the different electromagnetic nature of the
emitted particle (proton or neutron) in these transitions can in principle
be used to probe
and/or to emphasize different physics aspects. Of course, for experimental
tests, $^{37}$Ar may be preferred over $^{37}$K, since the latter is very
short--lived
while the former is a noble gas.

The nuclear structure calculations were
performed using the modern deformed mean field description provided by density
dependent Hartree--Fock (DDHF). The dependence on the nuclear structure of
response functions and asymmetries comes through the scalar and vector
momentum distributions for these transitions. We find that our DDHF
results describe nicely the main properties of the (oblate,
$K^\pi=\frac{1}{2}^+$) ground state bands of these nuclei, and thus
provide a reliable starting point for the calculation of the scalar and
vector momentum distributions for the transitions considered. The
kinematics and transitions are chosen to probe basically the odd--nucleon wave
function of the target nuclei. Our DDHF results provide a deformed
odd--nucleon orbital with important $s$-- and $d$--wave components. This
$\ell$--wave mixing is manifested in different characteristic ways in the
scalar and vector momentum distribution, as well as in the polarized and
unpolarized responses.

For unpolarized target only the
monopole component of the scalar momentum distribution contributes.
For polarized target the response functions are sensitive to the higher
multipoles of the scalar momentum distribution and to the vector
momentum distribution. We show that the latter (which can only be probed
when the beam is polarized) depends strongly on the terms of interference
between different $\ell$--wave components of the deformed bound wave function
of the struck nucleon, predominantly interference between $s$-- and
$d$--waves for the cases studied. We also show how the characteristic $sd$
interference shape is manifested in $R^{X'}$ responses for different
nucleon--detection plane and polarization direction. On the contrary, the
scalar momentum distributions, and hence the $R^X$ response functions,
depend mainly on the incoherent sum of $s$-- and $d$--wave contributions,
showing a characteristic double--peaked structure. The asymmetries
$P_\Sigma$ and $P_\Delta$ exhibit also a strong $p$--dependence due to the
complex structure of the deformed orbital. We show that the characteristic
shapes of response functions and asymmetries obtained for the deformed HF
orbital disappear in the spherical limit.
A comparison of PWIA and DWIA results
in the spherical limit is presented to show that standard FSI and
deformation effects are different in nature and manifest in a different
manner.

Although for quantitative comparison with experimental data, DWIA calculations
with the deformed model should be also performed, the results presented in
sections III and IV allow us to conclude that FSI effects will not mask
the effects of deformation discussed here. Hence, measuring response
functions and/or asymmetries for these transitions with polarized deformed
target and beam, would provide a unique tool to see the $\ell$--wave mixing
in deformed bound orbitals.

\vspace{5mm}
{\LARGE\bf Acknowledgements}
\vspace{3mm}

One of us (JAC) wishes to thank the members of CTP (MIT) for their
hospitality and, particularly, T. W. Donnelly for stimulating discussion.
This work has been supported in part by DGICYT (Spain) under
contract No. PB92/0021--C02--01. JMU is carrying out the work as a part of
a Community training project financed by the European Commission under
Contract ERBCHBICT 920185.

\newpage
\pagestyle{empty}

\begin{center}
{\bf Figure Captions}
\end{center}

\begin{itemize}
\item {\bf Fig. 1}: Single--nucleon responses ${\cal R}^L$, ${\cal R}^T$,
${\cal R}^{LT}$ and
${\cal R}^{TT}$ for proton (a) and neutron (b) emission. The choices of
kinematics and single--nucleon current are as discussed in the text.

\item {\bf Fig. 2}: Same as Fig.~1 for the single--nucleon responses
$\bd{\cal R}^{LT'}$ and $\bd{\cal R}^{T'}$. Not shown are the normal
components $n$, that are zero for the kinematics chosen ($\phi_N=0$).

\item {\bf Fig. 3}: Scalar (a), and vector longitudinal (b) and
transverse (d) momentum distributions for the transition to the first
$2^+$ state in the residual nucleus $^{36}$Ar with polarized $^{37}$K
target. Also
represented is the momentum distribution for the same transition with
unpolarized target (c). The dominant $\ell \ell'$ contributions are shown
by short--dashed ($\ell=\ell'=0$), dashed ($\ell=0, \ell'=2$), and
long--dashed ($\ell=\ell'=2$) lines.

\item {\bf Fig. 4}: Response functions for the reaction
$^{37}\mbox{K}(\vec{e}, e' p)^{36}$Ar($2_1^+$) with polarized target
(solid lines) and with unpolarized target (dashed lines). Polarization
direction parallel to $\bd{q}$ and coplanar kinematics ($\phi_N=0$,
$\bd{P}^\ast  \parallel \bd{e}_z$) have been chosen (see text).

\item {\bf Fig. 5}: Same as Fig.~4 for the reaction
$^{37}\mbox{Ar}(\vec{e}, e' n)^{36}$Ar($2_1^+$).

\item {\bf Fig. 6}: Response functions $R^{LT'}$ and $R^{T'}$ for
the same reaction as Fig.~4, but with polarization perpendicular to the
scattering plane and out--of--plane kinematics ($\phi_N=60$,
$\bd{P}^\ast  \parallel \bd{e}_y$). Also shown are the contributions from
$s$--wave (short--dashed line), $d$--wave (long--dashed line) and
interference between $s$ and $d$--waves (dashed line).

\item {\bf Fig. 7}: Asymmetries for the reactions
$^{37}\vec{\mbox{K}}(\vec{e}, e' p)^{36}$Ar($2_1^+$) (top) and
$^{37}\vec{\mbox{Ar}}(\vec{e}, e' n)^{36}$Ar($2_1^+$) (bottom) calculated
with the deformed HF orbital (solid line) and with the spherical orbital in
PWIA (dashed line) and DWIA (short--dashed line). The scattering angle is
$\theta_e=30$. Other kinematical variables are as in Fig.~4.

\item {\bf Fig. 8}: Response functions for
$^{37}\vec{\mbox{K}}(\vec{e}, e' p)^{36}$Ar in the spherical limit for
different polarization directions and $\phi_N$ values, calculated in PWIA
and in DWIA.
\end{itemize}

\newpage

\begin{table}
\begin{center}
\caption{Coefficients ${\cal F}^\lambda(\ell j;\ell' j')$ for the transition
from
the ground state ($J_A=\frac{3}{2}, K_A=\frac{1}{2}$) in $^{37}$Ar
($^{37}$K) to the first excited state ($J_B^\pi=2^+, K_B=0$) in $^{36}$Ar.}
\vspace{1.3cm}
\begin{tabular}{|c|l|cccc|}
  \hline
  &\,\,\,\,\,\,  $\ell_j$  &  $s_\frac{1}{2}$ &  $d_\frac{3}{2}$  &
                              $d_\frac{5}{2}$  &  $g_\frac{7}{2}$  \\
  & $ \ell'_{j'}$  &    &    &    &    \\ \hline \hline
  $\lambda=0$ &  $s_\frac{1}{2}$   & 0.080 &  0  &  0 &  0   \\
       & $d_\frac{3}{2}$ &  0  &   0.040   &  0  &  0        \\
       & $d_\frac{5}{2}$ &  0  &  0  &  0.011    &  0        \\
       & $g_\frac{7}{2}$ &  0  &  0  &  0  &  0.051          \\
                                                      \hline\hline
  $\lambda=2$ &  $s_\frac{1}{2}$   & 0 &  0.023  &  --0.028 &  0   \\
       & $d_\frac{3}{2}$ &  0.023  &   0.024   &  0.006  &  --0.029   \\
       & $d_\frac{5}{2}$ &  --0.028  &  0.006  &  0.001    &  --0.007  \\
       & $g_\frac{7}{2}$ &  0  &  --0.029  &  --0.007  &  0.037        \\
                  \hline
\end{tabular}
\end{center}
\end{table}

\begin{picture}(0,0)(0,0)
%\put(126,262.5){\line(6,-5){37}}
\end{picture}

\newpage

\begin{table}
\begin{center}
\caption{Coefficients ${\cal G}^L_l(\ell j;\ell' j')$ for the same transition
as
in table I.}
\vspace{1.3cm}
\begin{tabular}{|c|l|cccc|}
  \hline
  &\,\,\,\,\,\,  $\ell_j$  &  $s_\frac{1}{2}$ &  $d_\frac{3}{2}$  &
                              $d_\frac{5}{2}$  &  $g_\frac{7}{2}$  \\
  & $ \ell'_{j'}$  &    &    &    &    \\ \hline \hline
  $L=0$ & $s_\frac{1}{2}$ & --0.048 &     0     &  0     &  0        \\
        & $d_\frac{3}{2}$ &  0      &   --0.048 &  0.016 &  0        \\
        & $d_\frac{5}{2}$ &  0      &     0.016 &  0.006 &  0        \\
        & $g_\frac{7}{2}$ &  0      &     0     &  0     &--0.031    \\
                                                      \hline\hline
  $L=2$ & $s_\frac{1}{2}$ &  0     & --0.068 &  0.004    &  0        \\
        & $d_\frac{3}{2}$ &--0.068 &   0.010 &  0.010    &  0.004    \\
        & $d_\frac{5}{2}$ &  0.004 &   0.010 &$<10^{-3}$ &--0.018    \\
        & $g_\frac{7}{2}$ &  0     &   0.004 &--0.018    &  0.047    \\
                                                      \hline\hline
  $L=4$ & $s_\frac{1}{2}$ &  0      &     0     &  0     &  0.027    \\
        & $d_\frac{3}{2}$ &  0      &   --0.005 &  0.002 &--0.014    \\
        & $d_\frac{5}{2}$ &  0      &     0.002 &--0.001 &  0.005    \\
        & $g_\frac{7}{2}$ &  0.027  &   --0.014 &  0.005 &--0.008    \\
                                                      \hline
\end{tabular}
\end{center}
\end{table}
\begin{picture}(0,0)(0,0)
%\put(126,310){\line(6,-5){37}}
\end{picture}

\newpage

\begin{table}
\begin{center}
\caption{Coefficients ${\cal G}^L_t(\ell j;\ell' j')$ for the same transition
as
in table I.}
\vspace{1.3cm}
\begin{tabular}{|c|l|cccc|}
  \hline
  &\,\,\,\,\,\,  $\ell_j$  &  $s_\frac{1}{2}$ &  $d_\frac{3}{2}$  &
                              $d_\frac{5}{2}$  &  $g_\frac{7}{2}$  \\
  & $ \ell'_{j'}$  &    &    &    &    \\ \hline \hline
  $L=2$ & $s_\frac{1}{2}$ &  0      &   --0.068 &--0.003 &  0        \\
        & $d_\frac{3}{2}$ &--0.068  &     0.009 &  0.006 &--0.003    \\
        & $d_\frac{5}{2}$ &--0.003  &     0.006 &--0.005 &--0.022    \\
        & $g_\frac{7}{2}$ &  0      &   --0.003 &--0.022 &  0.042    \\
                                                      \hline\hline
  $L=4$ & $s_\frac{1}{2}$ &  0     &   0     &  0        &  0.014    \\
        & $d_\frac{3}{2}$ &  0     & --0.003 &  0.001    &--0.007    \\
        & $d_\frac{5}{2}$ &  0     &   0.001 &$<10^{-3}$ &  0.002    \\
        & $g_\frac{7}{2}$ &  0.014 & --0.007 &  0.002    &--0.004    \\
                                                      \hline
\end{tabular}
\end{center}
\end{table}
\begin{picture}(0,0)(0,0)
%\put(127,235.5){\line(6,-5){37}}
\end{picture}

\newpage

\begin{table}
\begin{center}
\caption{Weights of the $\ell j$--projection of the deformed odd--nucleon
wave function in $^{37}$Ar and $^{37}$K. Only values larger than
$10^{-3}$ have been tabulated.}
\vspace{1.3cm}
\begin{tabular}{cccccc}
\hline
\vspace{-4mm}
\\
$\ell_j$  & $s_\frac{1}{2}$  & $d_\frac{3}{2}$  & $d_\frac{5}{2}$
          & $g_\frac{7}{2}$  & $g_\frac{9}{2}$ \\
\vspace{-4mm} \\
\hline \hline
\vspace{-3mm}
\\
$^{37}$Ar  &  0.191  & 0.746  & 0.054  & 0.007 & 0.002 \\
$^{37}$K  &  0.194  & 0.741  & 0.056  & 0.007 & 0.002  \\
\vspace{-3mm}
\\
\hline\hline
\end{tabular}
\end{center}
\end{table}

\begin{picture}(0,0)(0,0)

\end{picture}
\newpage

\begin{table}
\begin{center}
\caption{Binding energies, charge r.m.s radii, ground state magnetic
moments, intrinsic quadrupole moments, moments of inertia and decoupling
parameters for $^{37}$Ar and $^{37}$K. Theoretical results
obtained from DDHF calculations are compared to available experimental
data\protect\ct{led}.}
\vspace{1.3cm}
\begin{tabular}{cccccccccc}
\hline
Nuclei  &
\multicolumn{2}{c} {$B$(MeV)} &
                    $r_c$(fm) &
\multicolumn{2}{c} {$\hspace{5mm} \mu_{3/2}$} &
                $Q_0$(fm$^2$) &
 ${\cal I}$(MeV$^{-1}$) & $a$
\\ \hline
  &  exp.  &  th.  & th.  & exp. &  th.
  &  th.   &  th.  & th.
\\ \hline
$^{37}$Ar & --315.50 & --312.28 & 3.41 & $.95\pm .20$ &
1.250 & --43.09 &  1.89  & --1.159\\
$^{37}$K & --308.57 & --305.88 & 3.43 & .203 &
.194 & --32.39 &  1.87 &  --1.140\\
\hline
\end{tabular}
\end{center}
\end{table}

\end{document}